 \documentclass[preprint,review,12pt]{elsarticle}
\usepackage{jasr}
\usepackage{framed,multirow}
\usepackage{float}
\usepackage{amssymb}
\usepackage{latexsym}
\usepackage{xcolor}
\usepackage{gensymb}
\usepackage{textcomp}

\usepackage{lineno}

\usepackage{url}
\usepackage{xcolor}
\definecolor{newcolor}{rgb}{.8,.349,.1}

\usepackage[citebordercolor=white]{hyperref}
\PassOptionsToPackage{unicode}{hyperref}
\PassOptionsToPackage{naturalnames}{hyperref}

\journal{Advances in Space Research}

\begin{document}

\verso{Deepthi Ayyagari \textit{et. al}}

\begin{frontmatter}

\title{Ionospheric response during Tropical Cyclones-a brief review on Amphan and Nisarga}%

 \author[1]{Deepthi \snm{Ayyagari}\corref{cor1}}  
\author[1]{Soumen \snm{Datta}}
\author[1]{Saurabh \snm{Das}}
\author[1]{Abhirup \snm{Datta}}
\address[1]{Department of Astronomy, Astrophysics and Space Engineering\unskip, Indian Institute of Technology Indore\unskip, Simrol \unskip, Indore\unskip, 453552, Madhya Pradesh, India}

{\cortext[cor1]{Corresponding author:Deepthi Ayyagari \ead{nagavijayadeepthi@gmail.com} Tel.: +91-996-316-6161;}}

% \received{1 May 2013}
% \finalform{10 May 2013}
% \accepted{13 May 2013}
% \availableonline{15 May 2013}
% \communicated{S. Sarkar}

\begin{abstract}
%%%
Here, we explore the different characteristics of a possible coupling between tropospheric and ionospheric activities during the impact of tropical cyclones (TC) like Amphan and Nisarga in the Indian subcontinent. We have analyzed the effect of TCs Amphan and Nisarga on the low latitude ionosphere using the measurements from several IGS stations around India and a GPS+ NavIC station in Indore, India. For the first time, this study assesses the impact of tropical cyclones on the equatorial ionosphere using both GPS and NavIC. After the landfall of TC Amphan, the VTEC analysis shows a significant drop from nominal values in both NavIC as well in GPS by $5.1$ TECU and $3.6$ TECU, respectively. In contrast to TC Amphan, Nisarga showed a rise in VTEC which ranged from $0.9$ TECU in GPS to $1.7$ - $5$ TECU in NavIC satellites except for PRN6. The paper examines Outgoing Longwave Radiation as a proxy to the convective activity which may be responsible for the ionospheric variation through the generation of gravity waves. In addition, the horizontal neutral wind observations at the location of TC landfall confirm the presence of ionospheric disturbances. VTEC perturbation analysis using a band-pass filter reveals a variation in differential TEC values between $\pm0.4$ and $\pm0.8$ based on the IGS station measurements. This indicates that the gravity wave is one of the responsible mechanisms for the lower-upper atmospheric coupling during both cyclones.
%%%%
\end{abstract}


\end{frontmatter}

%% For linenumbers
%\linenumbers
%\begin{linenumbers}

\section{Introduction} \label{sec1}

The last five decades have seen many advances in ionospheric physics centered around issues related to ionospheric irregularities, height variations along with the peak densities of various layers of ionosphere \citep{1}. The ionosphere is primarily impacted by electromagnetic radiation from space, seasonal changes in solar flux, and sudden events on the sun, such as coronal mass ejections (CME). It is these events that cause intense ionization in the ionosphere. GPS as well as the other ground-based radio systems are affected by this phenomenon, resulting in reduced operational reliability \citep{2,3}. According to earlier studies, the lower layers of the ionosphere are coupled with the neutral atmosphere. Neutral particle dynamics are dominant in the lower atmosphere, affecting the bottom-most layers of the ionosphere. The number of neutral particles is several orders greater in magnitude than the number of ionized particles in the lower ionosphere. Because of the strong coupling between ions and neutral particles in the lower layers of the ionosphere, the neutral particles collide with ions \citep{4}. Numerous experiments using sensitive instruments indicate that ionospheric disturbances are linked to atmospheric activity in the troposphere. Moreover, convective activities of the troposphere (tropical cyclones, hurricanes, typhoons, etc.) can play an integral role in triggering small and medium-scale ionospheric disturbances \citep{5,6,7}.

A chain of interconnected processes within the lithosphere-atmosphere-ionosphere interaction system reacts to various phenomena such as lightning discharge, high-power transmitters, high-power explosions, earthquakes, volcano eruptions, and cyclones\citep{48,49}. Through the global electric circuit (GEC), thunderstorms and cyclonic storms transfer energy from the atmosphere to the ionosphere and establish electricity between the atmosphere and ionosphere\citep{50}. Large-scale convection transports the charged aerosols and water droplets upward in a cloudy environment \citep{9,53}. \textbf{Large scale convection transports the charged aerosols and water droplets upward resulting in changes in electric circuit between the ground and ionosphere on a horizontal scale of hundreds of kilometers \citep{15,51,52,53}.} Furthermore, experimental observations and the simulation models revealed that atmospheric gravity wave (GW), which is a most common consequence of convective system like cyclone or thunderstorm, can propagate from the convective plumes in upward direction and can reach up to the ionospheric height.\citep{10,62,43,75,76}.
 
Tropical cyclones (TC) by definition are rotational low-pressure systems (classification presented in Table \ref{table:1}). They originate over the oceans in the equatorial region between 5\ensuremath{^{o}}and 20\ensuremath{^{o}} latitudes. In general, the central pressure of TC drops to a value of 5 to 6 hectopascals (hPa) and the maximum sustained wind speed exceeds 62 kilometers per hour (kmph). In general, TC is a powerful vortex structure, with a diameter of 150 km to 800 km by spiraling around a center along the surface of the sea associated with windspeed at the rate of 300 to 500 kmph \citep{8,54,55,57}. Hence a TC is described as an extending disturbance, having its effect spread over 1000–2000 km westward or eastward from the TC location. An in-depth study of cyclone Davina shows that the GWs produced by TCs can alter the composition of the Middle Atmosphere \citep{11}. Following the above studies, a significant disturbance in electron density measurement has been observed during the active phase of a TC in the D (60-80 km) region \citep{12,56}, F region\citep{77,78}  of the ionosphere. Within the effect zone of TC, ionospheric electron density increases and reaches a maximum before landfall of a TC, and decreases approximately a day after landfall\citep{13}. On the day following the landfall of a typhoon, observations from more than 50 GPS stations revealed (along the path of the typhoon Matsa) a gain of 5 TECU over its average monthly value and a drop of 1 TECU below the average monthly value \citep{14}. A comprehensive study from the Indian sector, for TC Mahasen and TC Hudhud, reveals an anomalous decrease of vertical TEC (VTEC) value from the monthly mean for TC Mahasen as well as TC Hudhud as 3.8 TECU and 2.1 TECU respectively on the day of the landfall \citep{15}. A significamt ionospheric perturbation has also been reported during another tropical cyclone over the indian region \citep{79}. The study also reveals that geomagnetic conditions should be quiet to investigate the passage of cyclones and their response in the ionosphere \citep{43,44,45,46,47}.

\begin{table*}
\caption{The Tropical Cyclone system (low pressure system)  over Indian region is classified based on the maximum sustained wind speed(\textit{V}) for three minute average value as well as the pressure(\textit{P}) associated with the system. The units for the wind speed is given in kilometre per hour (kmph) and the units for pressure is in hectopascals (hPa)}

\centering
\begin{tabular}{|l|l|l|l|}
 \hline
Category& Wind Speed (\textit{V})& Pressure (\textit{P}) \\
 \hline
Super Cyclonic Storm &$\geq$ 221 kmph & $>$ 65.5 hPa\\
Extremely Severe Cyclonic Storm & 166-220 kmph&15.5-65.5 hPa\\
Very Severe Cyclonic Storm &118-165 kmph & 8.5-15.5 hPa\\
Severe Cyclonic Storm& 89-117 kmph&4.5-8.5 hPa\\
Cyclonic Storm& 63-88 kmph&3-4.5 hPa\\
Deep depression& 51-62 kmph&1-3 hPa\\
Depression& 31-50 kmph&$\leq$1 hPa\\
 \hline
\end{tabular}
\label{table:1}
\end{table*}

Based on the earlier studies, this paper presents the details of the ionospheric response to two tropical cyclones, over the Indian region, i.e TC Amphan(super cyclonic storm) and TC Nisarga(severe cyclonic storm) which hit the east and west coast of the Indian peninsula. 
The TEC observations from a new constellation of Indian origin, known as NavIC (an acronym for NAVigation with Indian Constellation) is used along with the TEC estimates from the IGS stations to investigate the amount of local ionospheric perturbation caused by TCs. The major advantage that we have in using the data from NavIC is that it is a regional satellite navigation system, which is a constellation of seven satellites (a combination of Geostationary Earth Orbit (GEO) and Geosynchronous Orbit (GSO) satellites) in its space segment. NavIC satellites transmit signals in the $L5$ and $S1$ bands, with carrier frequencies of 1176.45 MHz and 2492.028 MHz, respectively, in a 24 MHz bandwidth. It is engineered to provide positional accuracy information to Indian users as well as a 1500 km radius around its boundary, which is defined by a rectangular grid spanning from 30\ensuremath{^{o}} S to 50\ensuremath{^{o}} N in latitude to 30\ensuremath{^{o}} E to 130\ensuremath{^{o}} E in longitude\citep{39}.
Unlike GPS, NavIC is designed to provide continuous spatial as well as temporal coverage (24 x 7) in the these regions. The reliability of NavIC in exploring the upper atmosphere is well demonstrated by \citep{34,35,36,38,40,41,42,68,69,70,71,72,73,74}. In general, the Waveforms derived from the ion density can demonstrate how the ionospheric plasma responds to the interaction with neutral gravity wave. Hence, a detrending of the ion density is done by calculating the temporal deviation, which gives us an idea of how the neutral wave-like gravity wave perturbs the ion density, and then we filter the deviation time series by a suitable bandpass filter in order to find the wave structure.

\section{Data and Methodology} \label{sec2}

A Global Navigation Satellite System (GNSS) receiver(Septentrio PolarX5s) along with Navigation with Indian Constellation (NavIC receiver: Accord Rx), provided by ISRO-Space Application Centre, is operational at the Space weather laboratory facility available at the Department of Astronomy, Astrophysics and Space Engineering, IIT Indore. These multi-constellations and multi-frequency receivers log in TEC data (as calculated from equation 1). Thus obtained TEC data from the receivers is used for the analysis to observe the ionospheric response of Tropical cyclones Amphan and Nisarga. 
The TEC estimates are the slant TEC values which depend on the length of the signal's path through the ionosphere and on the satellite elevation $E$ \citep{18,19,20,21,22,23}.

\begin{equation}
STEC = \int_{S}^{R}  {n_e} . {ds} 
\end{equation}

where $R$ is receiver, $S$ is the satellite, $n_e$ is the electron density, \textit{ds} is the path length. To redress this effect, an estimation of the Vertical TEC (VTEC) above a given point on the Earth's surface is essential. So VTEC is estimated with a single layer model mapping factor conversion, assuming all the free electrons are concentrated in a layer of infinitesimal thickness located at the altitude($I_h =$ 350 km) above the surface from the center of the earth($R_e =$ 6371 km) as shown in equations (2) and (3) \citep{24,25,26,27}
\begin{equation}
MF(E) = \frac{STEC}{VTEC}
\end{equation}
\begin{equation}
           MF(E)  = \Bigg[ \bigg[ 1 -  \Big[\frac{{R_e} * cos{(E)}}{R_e+I_h}\Big]^{2}\bigg]\Bigg]^{-1/2}
\end{equation}

\begin{figure}[H]
\centering
\includegraphics[width=2.5in,height=3in]{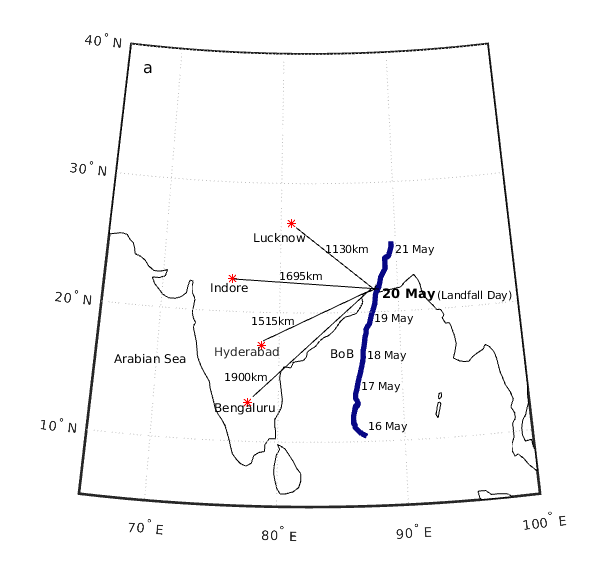}
\includegraphics[width=2.5in,height=3in]{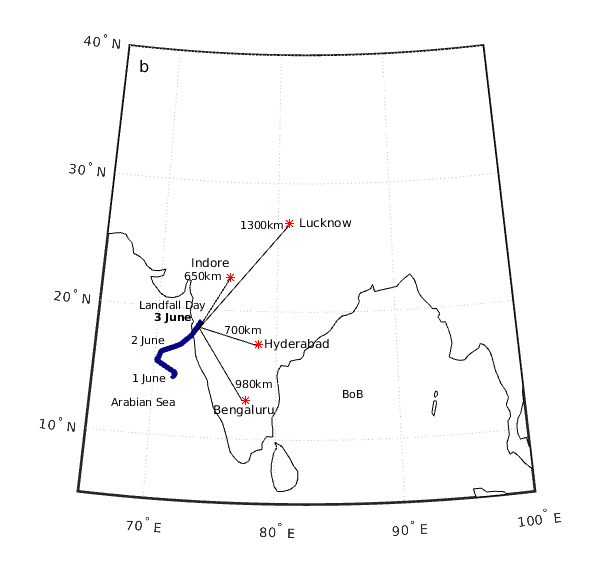}
    \caption{(a) TC Amphan trajectory(blue line) over the Bay of Bengal till the time of landfall from May 16 to 21, 2020. (b) TC Nisarga trajectory(blue line) over the Arabian Sea till the time of landfall from June 1 to 3, 2020. The red star in both the figures marks the location of the receivers used in this analysis and the arrow denotes the distance from the location of observation to the region of landfall}
\label{fig:figure 1}
\end{figure}

Apart from the VTEC measurements over the region Indore which is very near to northern crest of Equatorial Ionization Anomaly(EIA), the VTEC estimates from three different stations from the International Global Navigation Satellite System Service Network (IGS:\url{http://sopac-csrc.ucsd.edu/}) have also been used from 
(i) Lucknow, located far away from the northern crest of EIA, 
(ii) Hyderabad, located between the northern crest of EIA and the magnetic equator and 
(iii) Bengaluru, located near the magnetic equator.
Figure \ref{fig:figure 1} depicts the tracks of both TC along with the locations of the receivers and the distance from the landfall areas in the case of each tropical cyclone.

Table \ref{table:2} presents the geographic and magnetic dip location of receivers used in this study.

\begin{table*}
\caption{The Receiver Locations along with the Magnetic Dip locations used in the analysis are tabulated as follows.}
\centering
\begin{tabular}{|l|l|l|l|}
\hline
Station& G.Latitude &G.Longitude & Magnetic Dip \\
 \hline
Indore &22.52\ensuremath{^{o}}N&75.92\ensuremath{^{o}}E &32.23\ensuremath{^{o}}N \\
Lucknow&26.91\ensuremath{^{o}}N&80.95\ensuremath{^{o}}E &39.75\ensuremath{^{o}}N \\
Hyderabad&17.41\ensuremath{^{o}}N&78.55\ensuremath{^{o}}E &21.69\ensuremath{^{o}}N \\
Bengaluru&13.02\ensuremath{^{o}}N&77.57\ensuremath{^{o}}E &11.78\ensuremath{^{o}}N \\
\hline
\end{tabular}
\label{table:2}
\end{table*}

The gravity wave measurement has been computed based on the differential TEC estimates by detrending the differential carrier phase delay with a five minutes interval(\cite{31,32}): 
\begin{equation}
           s  = \Bigg[\frac{L_1 - L_2}{k} \Bigg] + b
\end{equation}
\begin{equation}
           \Delta^n s(t)  = \Delta^{n-1} s(t) -0.5\times \Bigg[\Delta^{n-1} s(t+\tau) + \Delta^{n-1} s(t-\tau) \Bigg]
\end{equation}
$n$ is the order of numerical difference; 
$L_1$ and $L_2$ are two frequency carrier phase measurements; 
$k$ is the factor for conversion to the VTEC measurement; $\tau$ is the time step which is considered to be five minutes here in the current study. 

After detrending, the signature of the ionospheric gravity wave has been measured by application of a band-pass filter with varying bandwidth of 1 to 3mHz (milli Hertz) \citep{33}. In addition, the outgoing longwave radiation (OLR) data is also used in the current analysis to probe for the energy density variation signatures. For tropical and subtropical regions, OLR values are used as proxy indicators of convection activity \citep{58}. Both top-of-the-atmosphere (TOA) and surface flux observations collected by the Clouds and the Earth's Radiant Energy System (CERES) sensors on NASA's Aqua and Terra satellites are used in this study \citep{16,17}. The daily average regional flux is estimated using diurnal models and the 0.25\ensuremath{^{o}} x 0.25\ensuremath{^{o}} regional fluxes at the hour of observation from the FLASHFlux monthly gridded TOA-surface flux. The data thus generated is space-time averaged data for OLR values during the whole period. Apart from OLR data the horizontal wind model (HWM14) data which is an empirical model of the horizontal neutral wind in the upper thermosphere is used. The updated model known as HWM14 consists of two parts: a quiet-time portion, and a geomagnetically disturbed portion that is dependent on the Ap index. The model currently is independent of solar activity as well as the F107 and F107A arguments. The HMW14 model provides zonal and meridional winds for specified latitude, longitude, time, and Ap index \citep{65}.

\section{Observational Results}\label{sec3}

\subsection{Ionospheric response to TC Amphan}
TC Amphan is the first tropical cyclone of 2020 in the North Indian Ocean basin, originated from a low-pressure area persisting a couple of hundred km east of Colombo, Sri Lanka around 6\ensuremath{^{o}}N. Tracking northeastward as shown in Fig.\ref{fig:figure 1}(a), the disturbance organized over exceptionally warm sea surface temperatures where TC Amphan underwent rapid intensification on May 17, 2020, and became an extremely severe cyclonic storm within 12 hours. On May 18, 2020, at approximately 12 UT(h), Amphan reached its peak intensity with three(3)-minute sustained wind speeds of 240 kmph and became the only super cyclonic storm in the last two decades over the Bay of Bengal. On May 20, 2020, between 10-11 UTC, the cyclone made landfall in West Bengal. At the time, the estimated Amphan's one-minute sustained winds were 155 kmph. The maximum sustained wind speed of any TC tends to fall during landfall. It further weakened and entered into a low-pressure area away from the state of West Bengal and moved away toward Bangladesh.

\begin{figure}[H]
\centering
\includegraphics[width=6in,height=5in]{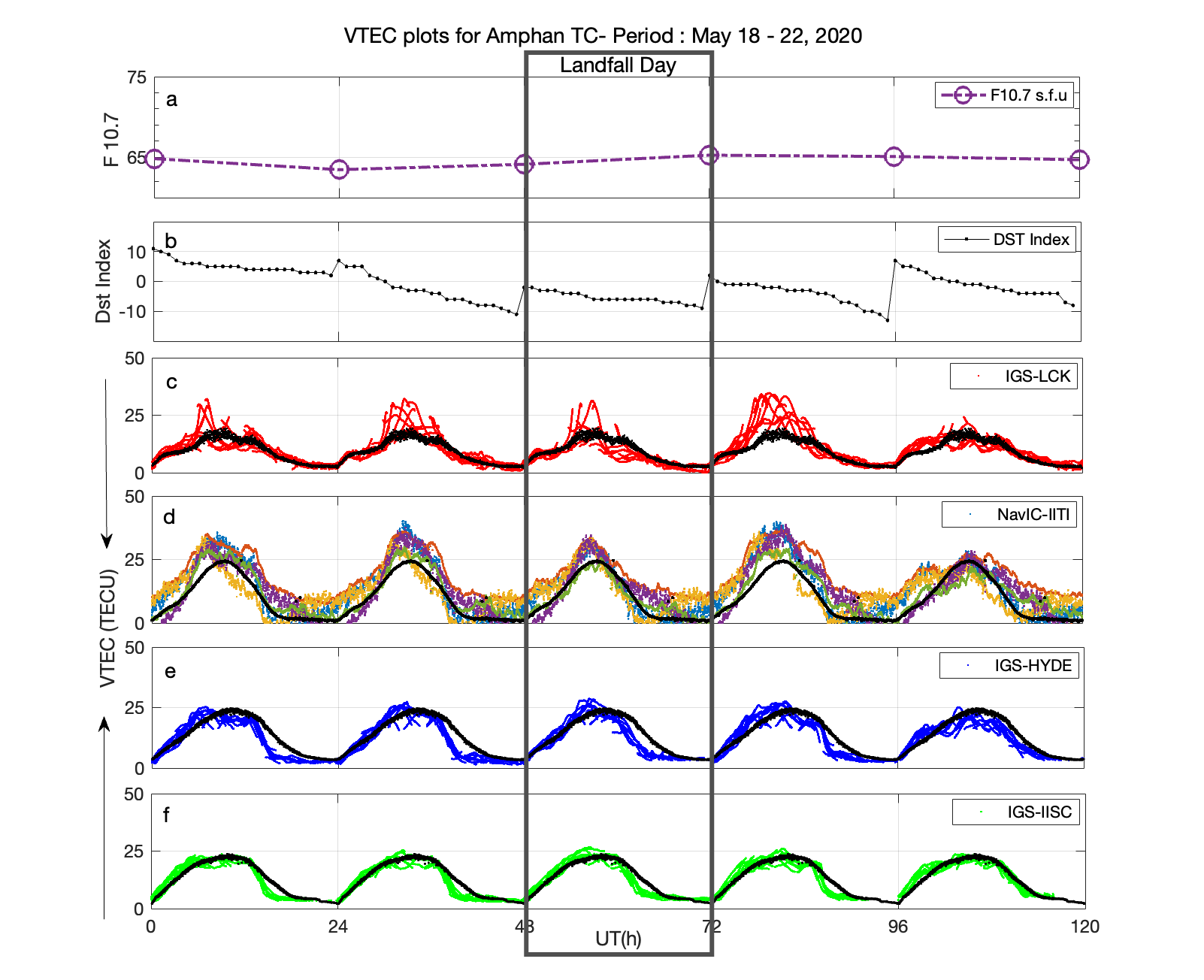}
    \caption{shows the VTEC analysis from May 18 to 22, 2020, for TC Amphan ionospheric response. (a) F10.7 cm Radio emissions (b) represent the Dst index (c)IGS-Lucknow (d)Indore-NavIC PRN(2 to 6) (e)IGS-Hyderabad (f)IGS-Bengaluru The black solid line in figures c-f represents the monthly mean of VTEC values.}
    \label{fig:figure 2}
\end{figure}

The Fig.\ref{fig:figure 2}(a) represents the F 10.7 solar flux units which indicates the solar activity for the TC Amphan period. Solar radio flux at 10.7 cm (2800 MHz) is a good indicator of solar activity. The F10.7 radio emissions are generated high in the chromosphere and low in the corona of the solar atmosphere. Unlike many solar indices, the F10.7 radio flux can easily be measured reliably on a day-to-day basis from the Earth’s surface, in all types of weather. Reported in solar flux units (s.f.u.), the F10.7 can vary from below 50 s.f.u. to above 300 s.f.u., over the course of a solar cycle. (b) Disturbance storm time (Dst) index which shows the geomagnetic activity during the days 18 to May 22, 2020. In general, to classify the severity of geomagnetic storms, the vital parameter is the Dst index, which measures the horizontal component of the Earth's magnetic field (H) in nano Tesla (nT). During such disturbances, this field gets depressed and its magnitude, which is axially symmetric in nature, varies with the time measured from the onset of a storm. Severity of geomagnetic storms can be classified as moderate storm (-50 nT $\leq$ Dst $<$ -100 nT) and intense storm (-100 nT $\leq$ Dst $<$ -200 nT) \citep{28,29}. Here in case of TC Amphan the Dst index value has not dropped below -13\textit{nT} which is highly favourable to probe the response of the ionosphere during TC activities. Hence the TEC values from May 18 to 22, 2020, have been analyzed to detect the anomalies on the day of landfall (May, 20, 2020) of Amphan.
Fig. \ref{fig:figure 2} (c-f) displays the VTEC values as estimated from four different stations, namely, Lucknow, Indore, Hyderabad, and Bengaluru, which fall within the 2000 km horizontal zone from the track of TC Amphan. The black solid line indicates the monthly mean VTEC values for May 2020 for each of the stations respectively.

\begin{figure}[H]
\centering
\includegraphics[width=5in,height=5in]{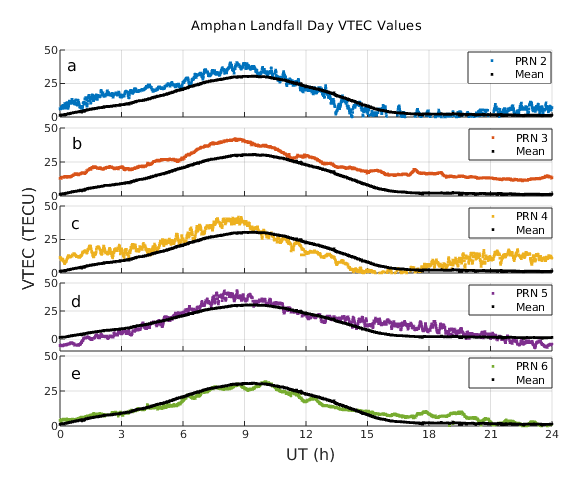}
    \caption{shows the VTEC analysis during May 20, 2020 for TC Amphan ionospheric response for NavIC PRNs 2 to 6(a-e). The black line represents the monthly mean of NavIC PRNs. }
    \label{fig:figure 3}
\end{figure}

On May 20, 2020, the VTEC values of all the available NavIC satellites identification numbers also known as Pseudo-Random Numbers (PRNs) for the Indore-NavIC receiver shown in Fig.\ref{fig:figure 3}(a-e). In this Fig.\ref{fig:figure 3} (a-e) the VTEC value is greater than the monthly mean value for all the PRNs. However, in the case of PRN6, the VTEC value is lower than its monthly mean value by 5.17 TECU. During the time of landfall i.e between 9 to 12 UT (hours) on May 20, 2020, PRNs 2, 3, 4, and 5 observed values higher than monthly mean values. The increased values above the monthly mean for each of the PRN are observed for PRN 2 above 1.11 TECU, PRN 3 above 4.43 TECU, PRN 4 above 2.34 TECU, and PRN 5 above 1.42 TECU respectively. On the contrary, after 12 UT (i.e 14:30 to 17:30 LT) decrease in TECU has been noted for all the 4 PRNs of NavIC which has not been reflected in the observations from PRN 3 that is depicted in Fig.\ref{fig:figure 3} (a-e). 
Likewise, a similar observation has been noted in the VTEC values from the IGS - reference stations Fig.\ref{fig:figure 2} (b,d,e) Lucknow (above 1.81TECU and below 3.64TECU), Hyderabad (above 0.63TECU and below 3.87TECU) and Bengaluru (above 1.73TECU and below 0.53TECU), before and after the time of landfall from the values of the monthly mean values respectively.

\begin{figure}[H]
\centering
\includegraphics[width=5in,height=6in]{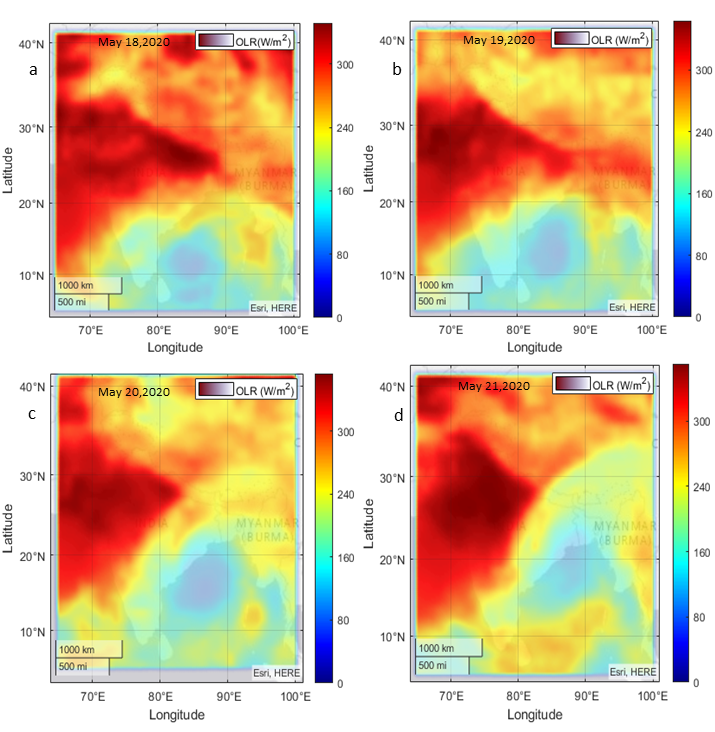}
    \caption{0.25\ensuremath{^{o}}x0.25\ensuremath{^{o}} grid CERES interpolated OLR data ($\frac{W}{m^{2}}$) between 10\ensuremath{^{o}}N to 40\ensuremath{^{o}}N Lat and 70\ensuremath{^{o}}E to 100\ensuremath{^{o}}E Longitude during Amphan period in 2020 for (a)18 May (b)19 May (c)20 May (d)21 May}
\label{fig:figure 4}
\end{figure}

The Fig.\ref{fig:figure 4}(a-d) shows the convective activity during the period of May 18 to 21, 2020, for TC Amphan. From the OLR surface interpolated data it is evident that there was one visibly increased activity between 10\ensuremath{^{o}}N to 30\ensuremath{^{o}} N Lat and 70\ensuremath{^{o}}E to 90\ensuremath{^{o}}E Lon on May 18, 2020, which has widespread region till May 19, 2020 up to 90\ensuremath{^{o}}E has merged abruptly during May 20, 2020, with decreased intensity and weakened on May 21, 2020. The energy values over the regions of Amphan trajectory grid is between 10\ensuremath{^{o}}N to 20\ensuremath{^{o}} N Lat and 80\ensuremath{^{o}}E to 90\ensuremath{^{o}}E Lon have varied from 240${W}/{m^{2}}$ on May 18, 2020 to 200${W}/{m^{2}}$ on May 19, 2020 and below 180${W}/{m^{2}}$ on May 20, 2020, and finally 160${W}/{m^{2}}$ on May 21,2020. However, the energy values between 20\ensuremath{^{o}}N to 30\ensuremath{^{o}} N Lat and 70\ensuremath{^{o}}E to 90\ensuremath{^{o}}E Lon, which is a (10x20)\ensuremath{^{o}} grid, remained uninterrupted with constant energy values above 300${W}/{m^{2}}$ before and after arrival as well as during the landfall of TC Amphan. Such low values of OLR during the landfall day are in agreement with earlier findings which clearly indicate the gravity wave energy of TCs is associated with low OLR values \citep{58,59,60}.

\subsubsection{Ionospheric response to TC Nisarga}

\begin{figure}[H]
\centering
\includegraphics[width=6in,height=5in]{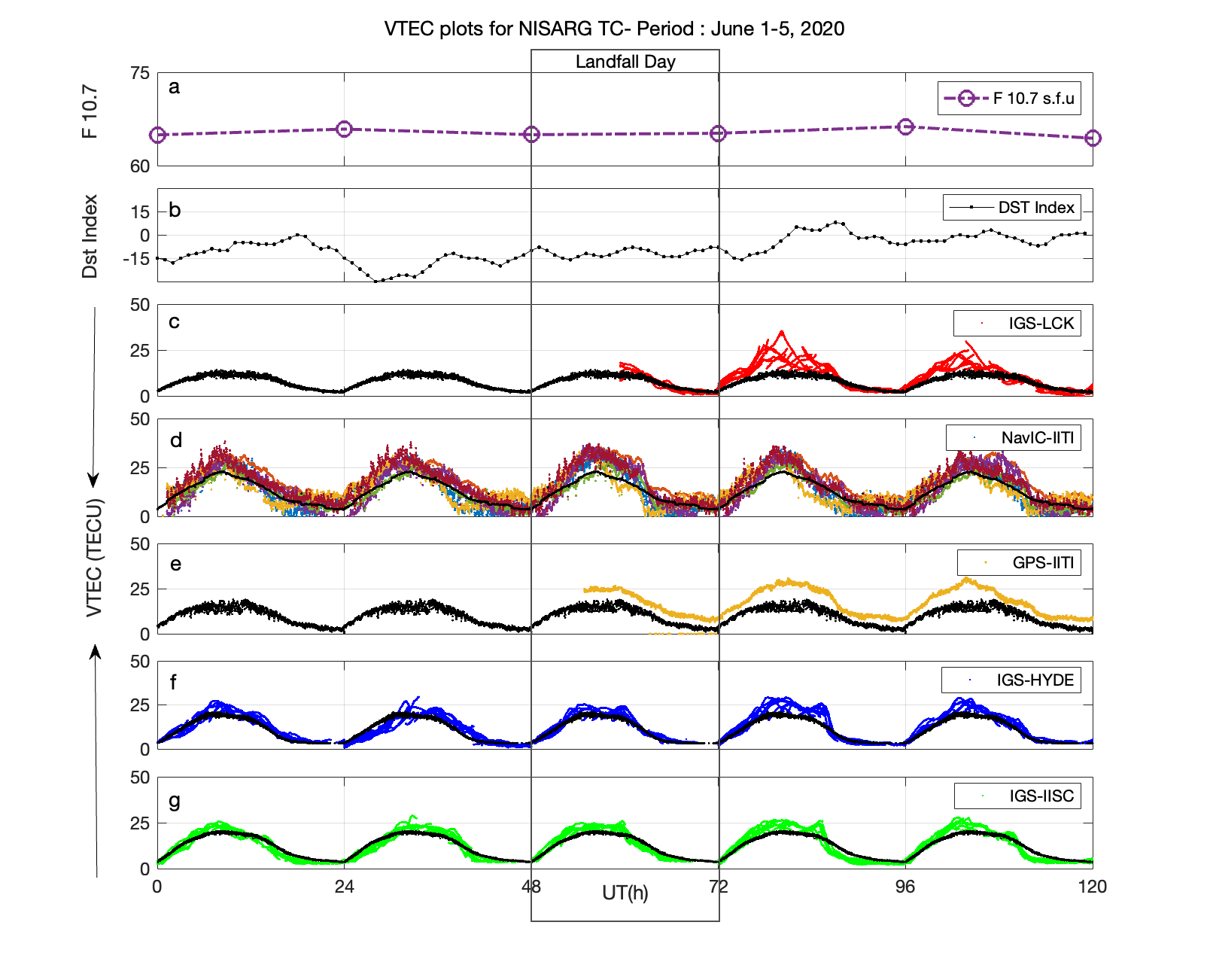}
    \caption{shows the VTEC analysis from June 1 to 5, 2020, for TC Nisarga ionospheric response. (a) F10.7 cm Radio emissions(b)the Dst index (c)IGS-Lucknow (d)Indore-NavIC PRN(2 to 6) (e)Indore-GPS (f)IGS-Hyderabad (g)IGS-Bengaluru. The black solid line in figures c-d-e-f-g represents the monthly mean of VTEC values.}
    \label{fig:figure 5}
\end{figure}

The second cyclone of the annual cyclone season, Nisarga originated as a depression in the Arabian Sea and moved gradually northward. In between 12-14 UT on June 2, 2020, the deep depression intensified into a cyclonic storm and thereby receiving the name Nisarga. It later intensified into the Deep Depression on the same day. TC Nisarga reached the peak intensity of 110 kmph which makes as a Severe Cyclonic Storm whereas a one-minute mean wind speed was 140 kmph which makes as a category 1 tropical cyclone. At 7 UT on June 3, 2020, Nisarga made landfall near the town of Alibag, at peak intensity which is in the vicinity of 700 km away from the Indore(NavIC) station Figure 1(b) represents the trajectory of TC Nisarga from 1 to June 3, 2020. The Fig.\ref{fig:figure 5} (a-g) presents the ionospheric response to TC Nisarga from June 1 to 5, 2020, for various stations as shown in Figure 1(b).
The Fig.\ref{fig:figure 5} (a) represents the F 10.7 solar flux units which indicate the solar activity for the TC Nisarga period. (b) shows the Dst index for this period which again has not dropped below the value of -30nT. Hence, TC conditions are much more favorable to record the ionospheric response during this period.

\begin{figure}[H]
\centering
\includegraphics[width=5in,height=4in]{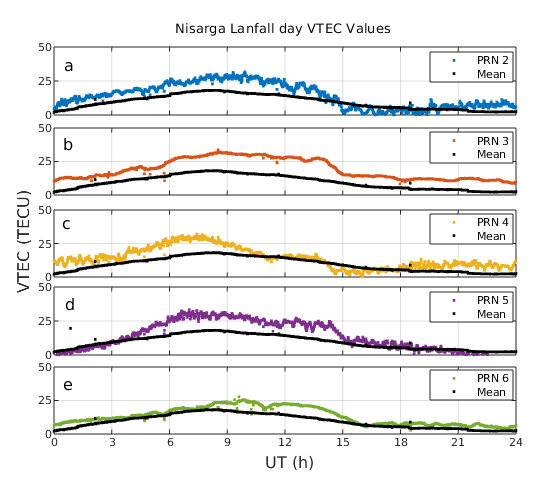}
    \caption{shows the VTEC analysis on June 3, 2020, for TC Nisarga ionospheric response as observed for individual NavIC PRNs 2 to 6(a-e).  The black solid line in the figures represents the monthly mean of VTEC values of NavIC PRNs.}
    \label{fig:figure 6}
\end{figure}

On June 3, 2020, the VTEC values of all the available 6 PRNs for the Indore-NavIC station (Fig.\ref{fig:figure 6}(a-e)), the TECU value is higher than the average of the mean of the month VTEC value except for PRN6( which is below 1.8TECU below the monthly mean value) during the time of landfall i.e between 6 to 9 UT(hours) when the maximum sustained wind speed reached the highest on June 3, 2020, whereas other reported values higher than monthly mean values, particularly during the time of landfall, PRN 2 above 3.92 TECU, PRN 3 above 4.88TECU, PRN 4 above 5.12TECU and PRN 5 above 1.74TECU respectively. Unlike TC Amphan before and after 6 to 9 UT (i.e 11:30 to 14:30 LT) increase of TECU has been noted on all 6 of PRNs.  A similar increase has been observed in the VTEC values as recorded by the IGS-reference stations(Fig.\ref{fig:figure 5}(c-e-f-g)), Hyderabad (above 0.9TECU), and Bengaluru (above 0.35TECU), before and after the time of landfall from the values of the monthly mean values respectively. The station Lucknow and Indore GPS have no VTEC measurements before and during the landfall time, but the mean values remained lower than the landfall day VTEC values.

\begin{figure}[H]
\centering
\includegraphics[width=5in,height=7in]{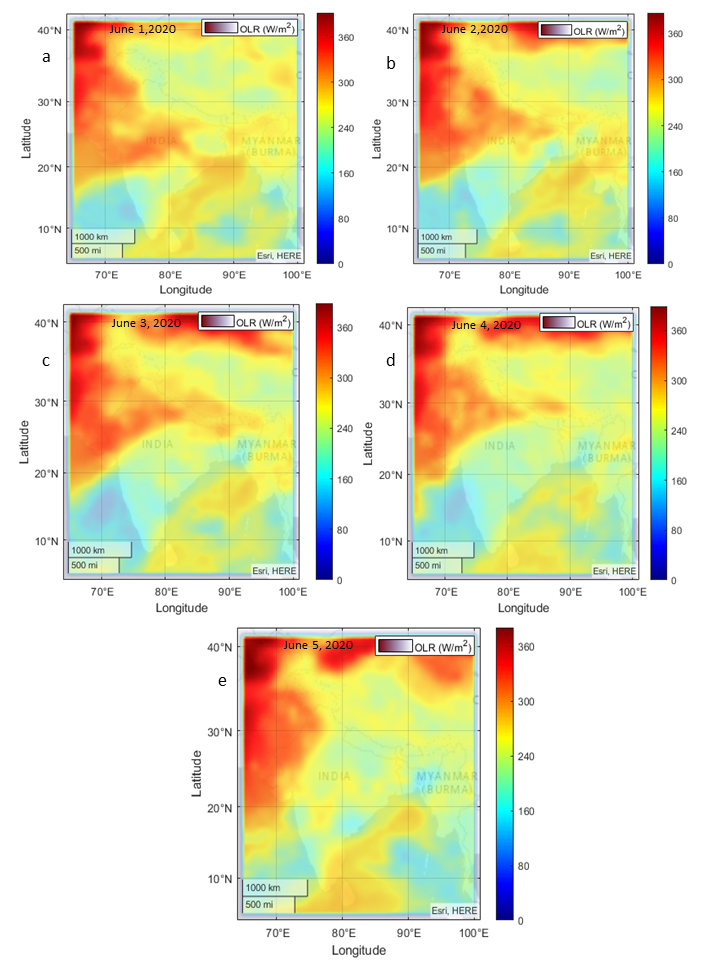}
\caption{0.25\ensuremath{^{o}}x0.25\ensuremath{^{o}} grid CERES interpolated OLR data ($\frac{W}{m^{2}}$) between 10\ensuremath{^{o}}N to 40\ensuremath{^{o}}N Lat and 70\ensuremath{^{o}}E to 100\ensuremath{^{o}}E Longitude during Nisarga,2020 for (a) 1 June (b) 2 June (c) 3 June (d) 4 June (e) 5 June}
\label{fig:figure 7}
\end{figure}

The Fig.\ref{fig:figure 7}(a-e) shows the convective activity during the period of June 1 to 5, 2020, for TC Nisarga. From the OLR surface interpolated data it is evident that the energy values between 10\ensuremath{^{o}}N to 30\ensuremath{^{o}} N Lat and 70\ensuremath{^{o}}E to 80\ensuremath{^{o}}E Lon on June 1, 2020, remained in the range of 260 to 280${W}/{m^{2}}$  till June 3, 2020, up to 85\ensuremath{^{o}}E has reduced below 260 ${W}/{m^{2}}$ on the landfall day of Nisarga. The energy values over the regions of the Nisarga trajectory grid that is between 10\ensuremath{^{o}}N to 20\ensuremath{^{o}} N Lat and 70\ensuremath{^{o}}E to 80\ensuremath{^{o}}E Lon have varied from 300${W}/{m^{2}}$ on June 1, 2020, to 240${W}/{m^{2}}$ June 4, 2020, finally revived back to 300${W}/{m^{2}}$ on June 5, 2020. The behavior of OLR values is consistent with that of TC Amphan and early findings. The OLR values remained to be lower than the surrounding OLR values especially in the region of TC Nisarga and during the landfall.

The above observations clearly show that for both the TCs, there was intense convective activity observed over the land region before and after the landfall of the TCs as well as when the TCs were in their active stages. The regions of merged convective activities during Amphan and Nisarga can be a probable explanation for the anomaly observed in VTEC values just after the landfall time on May 20 and June 3, 2020, respectively but the fact of rising of VTEC values during such intense convective activity can open up new studies in this field. This observation was in contrast to earlier observations made in the Indian Sector \citep{15}.

\subsection{Possible Mechanism: Gravity Wave Signature}

\begin{figure}[H]
\centering
\includegraphics[width=6in,height=4.2in]{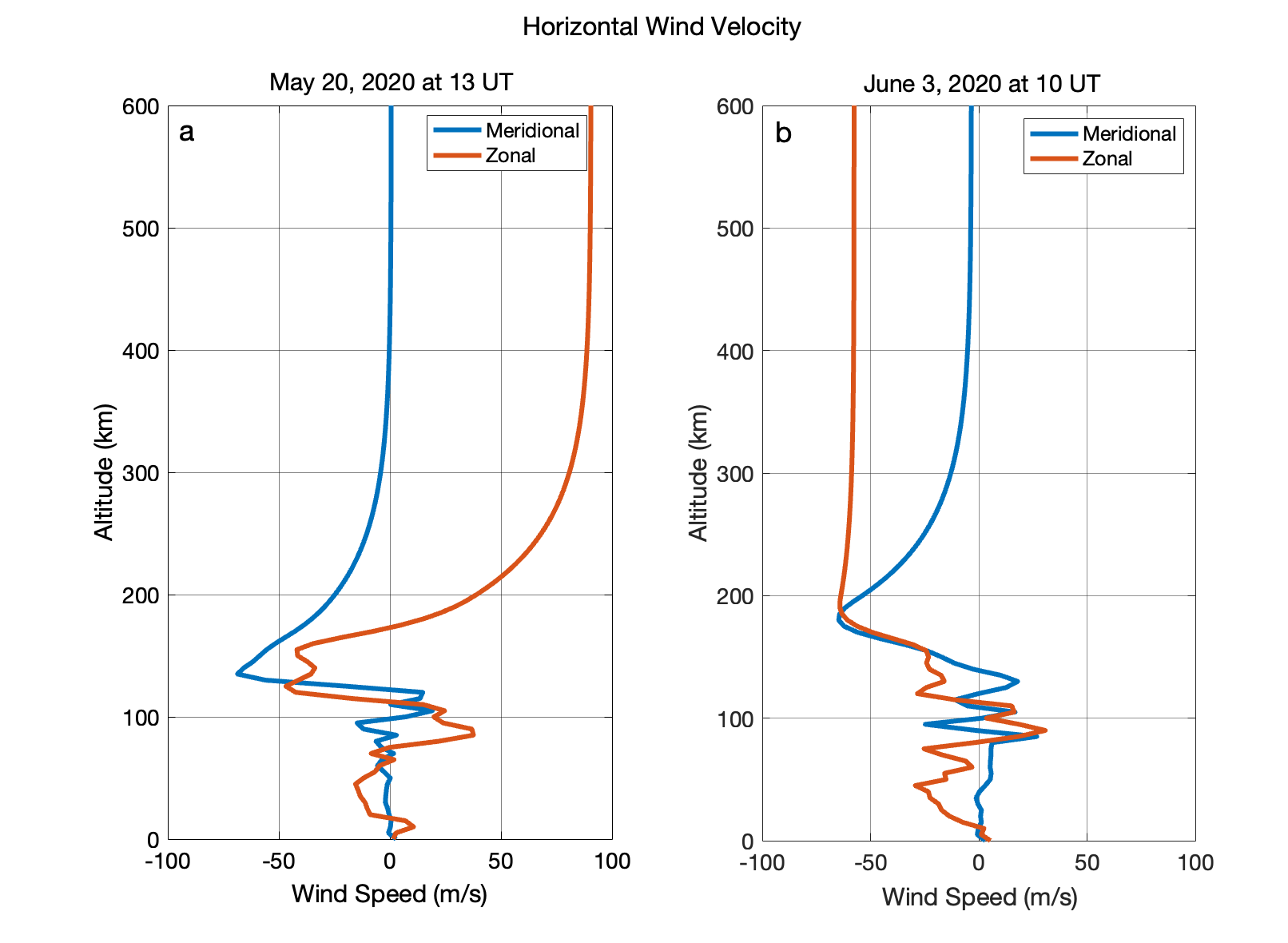}
    \caption{The variation of atmospheric Horizontal Wind Velocity with altitude as measured at locations: (a) Bakkhali, West Bengal (21.56\ensuremath{^{o}}N Lat. and 88.25\ensuremath{^{o}}E Lon. at 13 UT); (b) Alibaug, Maharashtra (18.65\ensuremath{^{o}}N Lat. and 72.87\ensuremath{^{o}}E Lon. at 10 UT)}
    \label{fig:figure 8}
\end{figure}

It is noteworthy to mention that horizontal neutral wind and temperature play a key role in the propagation of the gravity wave through the atmosphere \citep{61,62,63,64}. The horizontal neutral wind in the upper thermosphere has been defined using Horizontal Wind Model 14 (HWM14) an empirical model \citep{65}. The signature of the background atmospheric wind observations on May 20 and June 3, 2020, is quite evident. The variation of atmospheric horizontal wind velocity with altitude as measured at locations: Bakkhali, West Bengal, and Alibaug, Maharashtra is shown in Figure \ref{fig:figure 8} (a) \& (b).

A strong eastward wind has been observed above an altitude of 175Km which was along the northeast direction one hour before and towards the southeast direction from one hour after (not shown in Figure) on May 20. Hence the ionospheric perturbation or the traveling ionospheric disturbance signature towards the west direction (opposite to the horizontal wind) will be evident as compared to eastward. The statistical investigation \citep{80} and theoretical studies \citep{81,66} reported that gravity wave mostly propagates roughly against the neutral wind. It happens because gravity waves propagating along the wind have larger vertical wavelength as compared to gravity waves propagating against the wind. It results dissipation from viscosity at lower thermospheric altitude and critical level of filtering\citep{66,67,64}. We could not investigate the signatures of GW in multiple direction due to limited number of navigation receivers and it is not also within the scope of this study and would be incorporated for future study.

\begin{figure}[H]
\centering
\includegraphics[width=5in,height=2.8in]{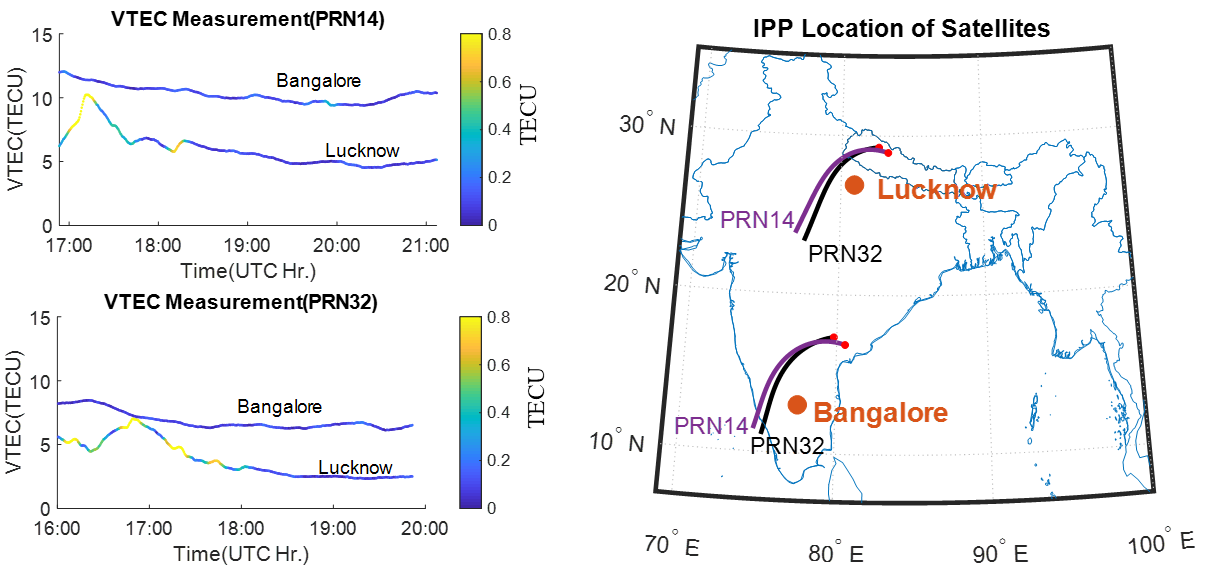}
    \caption{VTEC estimates from two different satellites observed from two different IGS stations on May 20, 2020.}
    \label{fig:figure 9}
\end{figure}

A detailed investigation have been done based on the perturbed TEC estimate during the TC period of Amphan and Nisarga. The preliminary analysis shows ionospheric disturbances after the TC Amphan. The TEC estimates were measured from two IGS stations Lucknow and Bengaluru. During the analysis, a sudden perturbed signature from the TEC estimates of IGS Lucknow station is observed and the variation of absolute VTEC with the signature of perturbation from two different satellite measurements on May 20, 2020, is shown in Fig.\ref{fig:figure 9}.The color bar shows the range of the perturbation amplitude in the estimates of the absolute VTEC estimates. However, the signatures from Bengaluru do not show any such anomaly.
\begin{figure}[H]
\centering
\includegraphics[width=6in,height=3in]{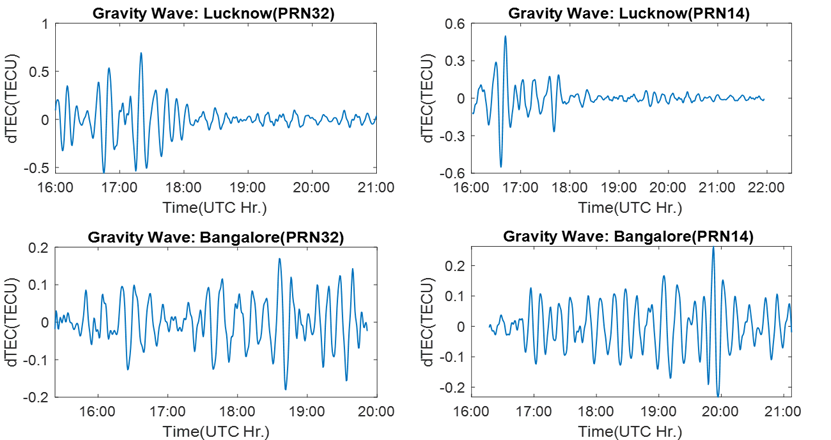}
    \caption{Gravity wave signature from the TEC estimates of four satellites as observed from IGS station Lucknow on May 20, 2020.}
    \label{fig:figure 10}
\end{figure}

\begin{figure}[H]
\centering
\includegraphics[width=5in,height=3in]{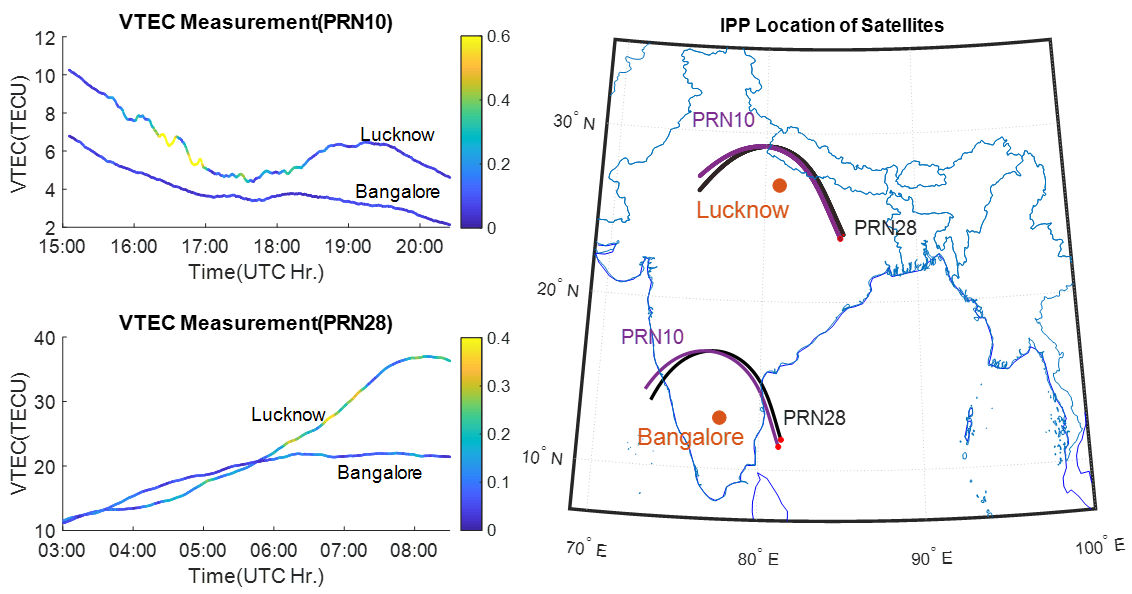}
    \caption{VTEC estimates from two different satellites as observed from two IGS stations on May 21, 2020.}
    \label{fig:figure 11}
\end{figure}

\begin{figure}[H]
\centering
\includegraphics[width=6in,height=3in]{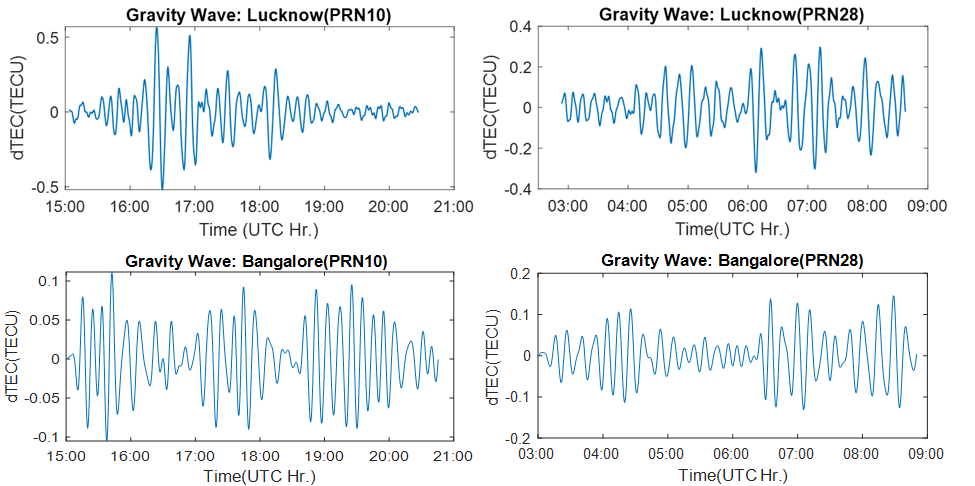}
    \caption{Gravity wave measurements from different satellite signal as observed from IGS station Lucknow and Bengaluru on May 21, 2020.}
    \label{fig:figure 12}
\end{figure}

It is fascinating to note that the landfall event of the TC Amphan began from 11 UT(h) on May 20, 2020, and the movement of the convective system along with the IPP location of satellite trajectories indicate the possible motion of neutral particles to the lower ionosphere. The retrieved GW measurements from different receivers by applying the band-pass filter, have also very significant signatures. The signal amplitude observed is above the value of $\pm0.1$ TECU which reaches up to 0.4 TECU to 0.6 TECU for the measurements from Lucknow station and 0.3TECU for Bengaluru station (Fig.\ref{fig:figure 10}). The dominant frequency for every cases obtained of about $2.0$\,mHz. It indicates the possible role of gravity wave behind the generation of this ionospheric perturbation signature. The same is in accordance with absolute VTEC signatures of ionospheric perturbations measured from Bengaluru and Lucknow IGS stations where the signatures obtained from Lucknow station are more evident when compared to Bengaluru station. The nature of perturbation in the local ionosphere is quite evident from two of the satellite observations on May 21, 2020. The variation of VTEC measurement with its differential measurement from IGS stations Lucknow and Bengaluru on this day is shown in Fig. \ref{fig:figure 11} The gravity wave signature is significant for this day also where the amplitude variation reached up to $\pm$ 0.4 TECU for both the stations. (Fig.\ref{fig:figure 12}).


\begin{figure}[H]
\centering
\includegraphics[width=5in,height=3.2in]{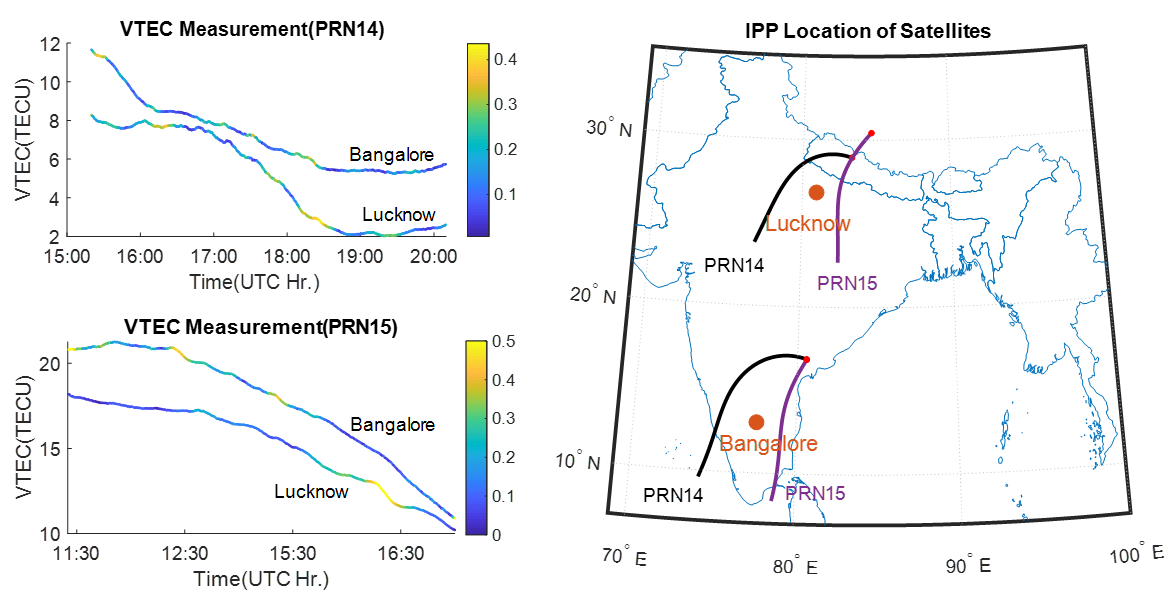}
    \caption{VTEC measurement from four different satellites observed from Lucknow and Bengaluru station on June 3, 2020.}
    \label{fig:figure 13}
\end{figure}

Later, as a follow-up to the TC Amphan analysis, a similar analysis has been performed with the TEC estimates during the TC Nisarga period from the data extracted from IGS Lucknow and Bengaluru stations. Fig.\ref{fig:figure 13} shows the absolute VTEC measurement with perturbation signatures from two different satellites on June 3, 2020, during different time on this day. The variations of the color in Fig.\ref{fig:figure 13} indicate the amplitude of perturbation.

\begin{figure}[H]
\centering
\includegraphics[width=6in,height=3in]{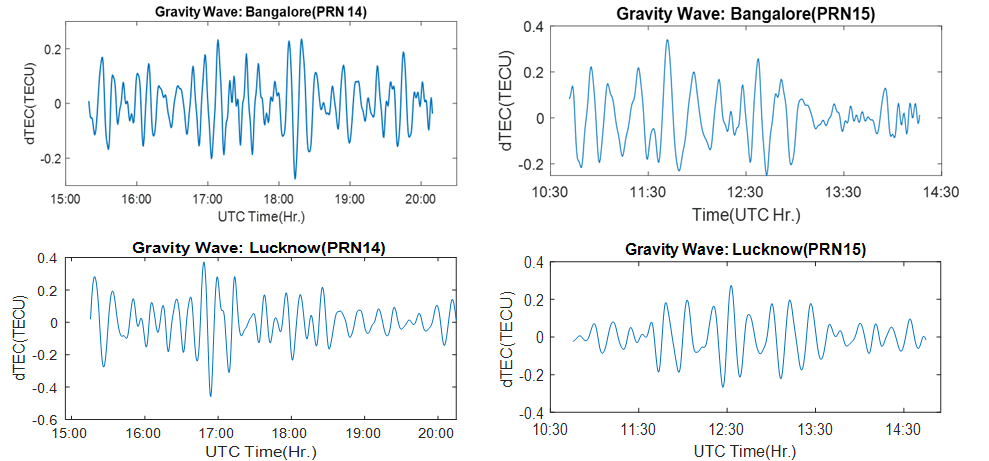}
    \caption{Gravity wave measurement from two different satellites observed from Bengaluru and Lucknow stations on June 3, 2020.}
     \label{fig:figure 14}
\end{figure}
 
\begin{figure}[H]
\centering
\includegraphics[width=5in,height=2.8in]{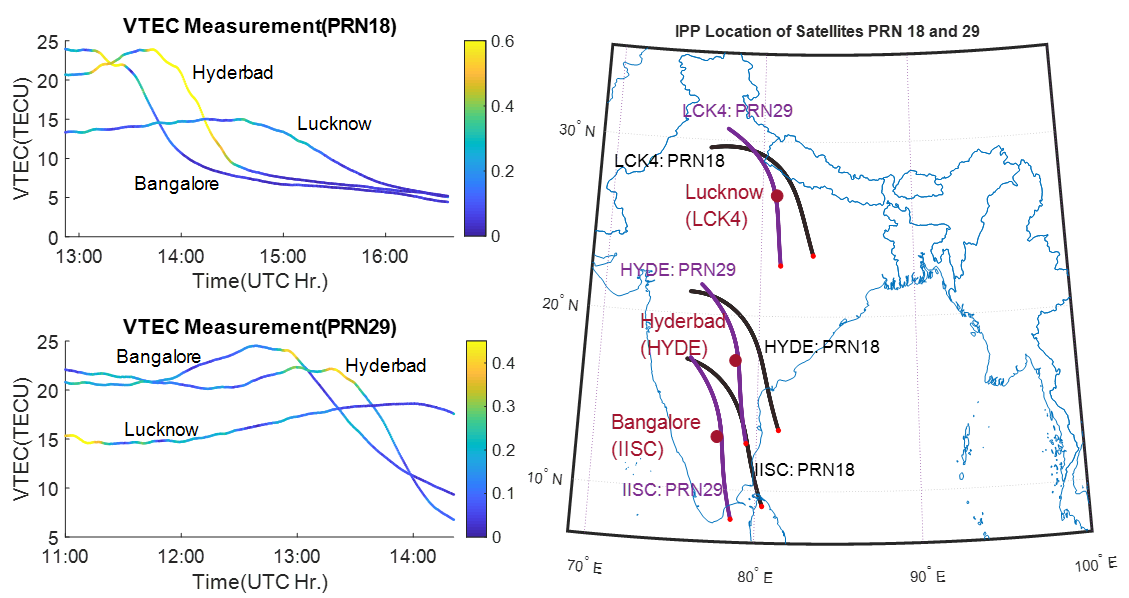}
    \caption{VTEC measurements from two different satellites observed from three IGS stations on June 4, 2020.}
    \label{fig:figure 15}
\end{figure}

Gravity wave signature based on differential VTEC measurement is shown in Fig.\ref{fig:figure 14} The observation states that a significantly low amplitude was detected from both the Bengaluru and Lucknow stations TEC analysis. A strong westward zonal wind from 100km altitude has been observed from the wind density variation at Alibaug, Maharashtra (18.65\ensuremath{^{o}}N Lat. and 72.87\ensuremath{^{o}}E Lon. at 10 UT) during 10 UT as observed in Fig. \ref{fig:figure 8} (b). Even-though, a sharp peak in the southward meridional wind has been observed around 180km altitude but the combined variation was westward for higher altitudes. Hence the significant ionospheric perturbance signature is expected to be observed along the eastward direction due to the propagation of gravity waves against the wind direction. The possible reason for the low amplitude signature in IGS Bengaluru and Lucknow is the long range of observation area from the possible source.The VTEC observation from three IGS stations (Bengaluru, Lucknow, and Hyderabad) on June 4, 2020, is observed as shown in Fig.\ref{fig:figure 15}. It is quite indicative that a different range of perturbations has been observed from various stations.
\begin{figure}[H]
\centering
\includegraphics[width=5in,height=3.8in]{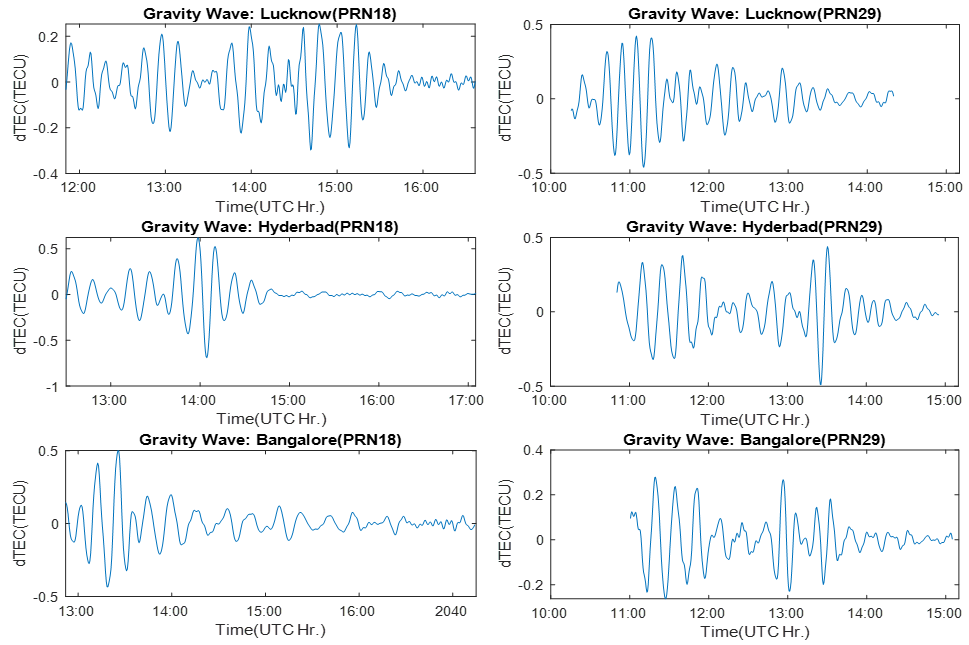}
    \caption{Gravity Wave measurements from two different satellites observed from two IGS stations on June 4, 2020.}
    \label{fig:figure 16}
\end{figure}
The variations observed at Hyderabad station are higher when compared to the other two stations. The signal waveform, as an outcome of the bandpass filter, also indicate the same and is shown in Fig.\ref{fig:figure 16}. A higher amplitude variation has been observed from the Hyderabad station which reached up to 0.6 TECU. The results obtained from IGS Lucknow and Bengaluru are also similar for this day. The amplitude of the perturbation has reached 0.4 TECU for the observations both from IGS Lucknow and Bengaluru stations. These different signatures obtained from different stations' observations can be attributed to the possible effect of wind flow. The IPP position of the satellite trajectories is directly opposite to the direction of zonal wind for the Hyderabad station and it is closer to the cyclone strike area. Hence the dissipation from viscosity will be less at a lower thermospheric altitude around IGS Hyderabad station and the effect is more for the other two stations also which has been reflected in our observation. A continuous six days(June 1 to 6, 2020) measurement of gravity wave by two different satellites (PRN 25, 28, and 29) from Bengaluru station has been shown in Fig. \ref{fig:figure 17}. It has been observed from all satellite measurements that a higher perturbation occurred on June 3 and 4, 2020, and gradually decreased in the next two days. Due to the unavailability of data from June 1-3, 2020, the investigation on these days can't be carried out from Hyderabad station observations. However, the signature from the Bengaluru station indicates the possibility of the generation of gravity waves that perturbed the ionosphere after the landfall of the cyclone.
\begin{figure}[H]
\centering
\includegraphics[width=5in,height=3.8in]{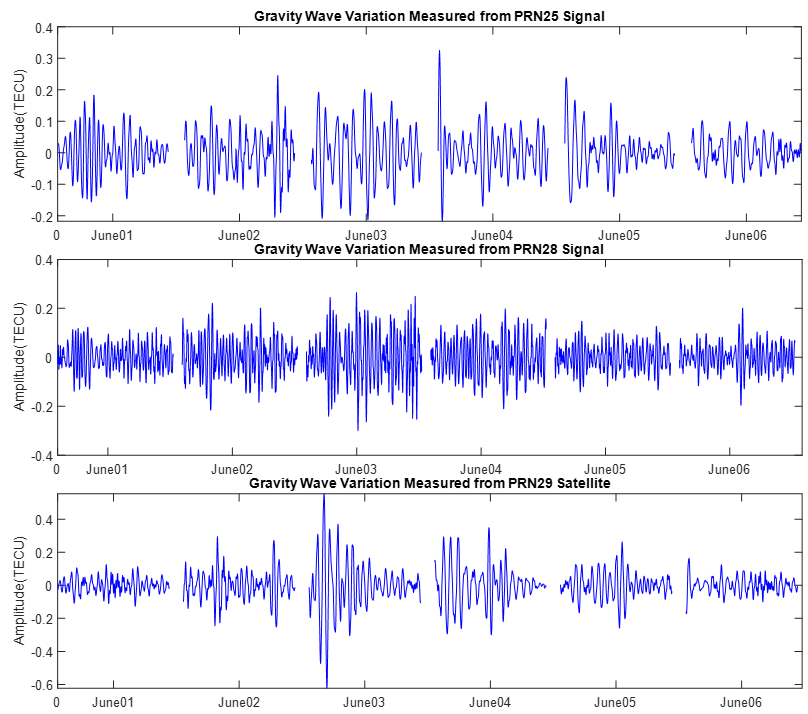}
    \caption{Gravity wave variation between June 1 to 6, 2020, from Bengaluru station as observed from three different satellite measurements.}
    \label{fig:figure 17}
\end{figure}

\section{Discussions and Conclusions} \label{sec4}

The analysis highlights the observation using NavIC, which is one of the recent regional navigation satellite systems launched specifically for the Indian subcontinent, on the effects of tropical cyclones Amphan and Nisarga on the equatorial ionosphere. As all the observations are registered at low geomagnetic conditions, they would aid in the detection of ionospheric perturbations in response to large-scale tropospheric activities. The results showed a sharp increase in VTEC for all the stations during the time of landfall and the values increased further on the next day of landfall in both the TC cases. These observations are contrary to earlier observations made by \citep{13,15} but in agreement with \citep{14}. However the OLR signatures were in accordance with earlier studies during both the cyclones \citep{10,30}.  In general, the rise of VTEC values above the monthly mean on the day of the landfall of the TCs has been observed. This observation, to some extent, is qualitatively in agreement with previous literature \citep{14}. As far as the disturbances produced by the landfall of TCs are concerned, the magnitude of VTEC increment produced by the TCs over the Bay of Bengal is much higher than the previously published results. These are some new observations that are the first of their kind from the Indian sector using a combination of NavIC and GPS.

To understand the possible mechanism and the level of ionospheric perturbation the horizontal meridional wind model data is utilized in this analysis. Based on the signatures of the HMW14 the gravity wave signature is explored for individual satellites. The waveform retrieved from the ion density was used in this study to observe the response of the ionospheric plasma to the interaction of neutral gravity waves. The variation of absolute VTEC along with the perturbation signature from different satellite measurements from Lucknow which are above 0.4 TECU, indicate the presence of gravity wave during TC Amphan. Likewise, significant signatures have been obtained for the case of TC Nisarga also. In both cases the role of zonal and meridional winds for the propagation of gravity wave has been found. Though a significantly low amplitude signal above 0.2 TECU has been detected from the Bengaluru station which is indicative of the far-field observation and the depletion of the perturbation signature against the meridional wind direction for TC Nisarga. These are clear signatures of gravity wave-induced perturbation of the ionosphere during these two cyclones.It is however to be noted that these preliminary results are affirmative and indicate the possible effect of cyclones on the ionosphere. However the study presented here is with the limited number of GPS stations. Hence, it is not possible to completely separate the effect of other probable sources which may also affect the ionosphere. Furthermore to understand such possible mechanisms data from more number of such observations around a TC-affected region in the vicinity (of within and beyond the 800-1000kms range) can be an extension to this study.

%\end{linenumbers}

\section*{Acknowledgments}
DA acknowledges the Department of Science and Technology for providing her with the INSPIRE fellowship grant to pursue her research. SAC, ISRO is further acknowledged by the authors for providing the NavIC receiver (ACCORD) under NGP-17 to the Department of Astronomy, Astrophysics and Space Engineering, IIT Indore. The authors would also like to acknowledge Prof. Gopi Seemala of the Indian Institute of Geomagnetism (IIG), Navi Mumbai, India for providing the software(\url{https://drive.google.com/file/d/1XgwY8iBtoHvqz8IVc4J8dGirgTs5BVwL/view?usp=sharing}) to analyze the IGS data( \url{http://sopac-csrc.ucsd.edu/index.php/data-download/}). Further acknowledgements go to the World Data Center for Geomagnetism, Kyoto for the Dst index data accessible via \url{http://wdc.kugi.kyoto-u.ac.jp/kp/index.html} and the Space Weather prediction Center (SWPC) under National Oceanic and Atmospheric Admisnistration for the F10.7 data archives accessible via \url{https://lasp.colorado.edu/lisird/}. In addition the authors also thank the NASA's CCMC for the HMW 14 model data accessible via \url{https://kauai.ccmc.gsfc.nasa.gov/instantrun/hwm}.

\newpage
%% Bibliography
%% Author year style
\bibliographystyle{jasr-model5-names}
\biboptions{authoryear}
\bibliography{refs}

\end{document}